\documentclass[letterpaper]{article} 
\usepackage{aaai2026}  
\usepackage{times}  
\usepackage{helvet}  
\usepackage{courier}  
\usepackage[hyphens]{url}  
\usepackage{graphicx} 
\def\UrlFont{\rm}  
\usepackage{natbib}  
\usepackage{caption} 
\usepackage{algorithm}
\usepackage{algorithmic}

\usepackage{newfloat}
\usepackage{listings}
\DeclareCaptionStyle{ruled}{labelfont=normalfont,labelsep=colon,strut=off} 
\floatstyle{ruled}
\newfloat{listing}{tb}{lst}{}
\floatname{listing}{Listing}

\usepackage[most]{tcolorbox}
\tcbuselibrary{breakable} 
\usepackage{graphicx}
\usepackage{amsmath}
\usepackage{amssymb}
\usepackage{enumitem}
\usepackage{array}
\usepackage{arydshln}
\usepackage{multirow}
\usepackage{microtype}
\usepackage{graphicx}
\usepackage{subfigure}
\usepackage{booktabs} 
\usepackage{xcolor}
\usepackage{algorithm}
\usepackage{algorithmic}
\usepackage{float}
\usepackage{listings}

\title{AgentTutor: Empowering Personalized Learning with \\ Multi-Turn Interactive Teaching in Intelligent Education Systems}
\author{
    Yuxin Liu\textsuperscript{\rm 1}\textsuperscript{\rm 2}\equalcontrib,
    Zeqing Song\textsuperscript{\rm 1}\textsuperscript{\rm 2}\equalcontrib,
    Jiong Lou\textsuperscript{\rm 1}\textsuperscript{\rm 3}, Chentao Wu\textsuperscript{\rm 1}\textsuperscript{\rm 3}, Jie Li\textsuperscript{\rm 1}\textsuperscript{\rm 3}\thanks{Corresponding Author}
}
\affiliations{
    \textsuperscript{\rm 1} School of Computer Science, State Key Laboratory of Submarine Geoscience, Shanghai Jiao Tong University \\ 
    \textsuperscript{\rm 2} SJTU Paris Elite Institute of Technology, Shanghai Jiao Tong University\\
   \textsuperscript{\rm 3} Shanghai Key Laboratory of Trusted Data Circulation and Governance and Web3\\
   Shanghai 200040, China

    \{emotion-xin, ignotus, lj1994, wuct, lijiecs\}@sjtu.edu.cn
}

\usepackage{bibentry}

\begin{document}

\maketitle

\begin{abstract}
The rapid advancement of large-scale language models (LLMs) has shown their potential to transform intelligent education systems (IESs) through automated teaching and learning support applications. However, current IESs often rely on single-turn static question-answering, which fails to assess learners' cognitive levels, cannot adjust teaching strategies based on real-time feedback, and is limited to providing simple one-off responses. To address these issues, we introduce AgentTutor, a multi-turn interactive intelligent education system to empower personalized learning. It features an LLM-powered generative multi-agent system and a learner-specific personalized learning profile environment that dynamically optimizes and delivers teaching strategies based on learners' learning status, personalized goals, learning preferences, and multimodal study materials. It includes five key modules: curriculum decomposition, learner assessment, dynamic strategy, teaching reflection, and knowledge \& experience memory. We conducted extensive experiments on multiple benchmark datasets, AgentTutor significantly enhances learners' performance while demonstrating strong effectiveness in multi-turn interactions and competitiveness in teaching quality among other baselines. 

\end{abstract}

\section{Introduction}
The development of large-scale language models (LLMs) has demonstrated outstanding capabilities in various natural language tasks \cite{wei2022chain, wambsganss2021arguetutor,xiong2024building, macina2023mathdial,zhang2024codeagent, md2024mapcoder,huang2024enhancing}. Among these, LLMs' exceptional zero-shot reasoning and interactive abilities hold great promise in the field of education \cite{jakub2023opportunities, kasneci2023chatgpt,jurenka2024towards,tack2022ai}. Leveraging these advantages, researchers have developed LLM-powered intelligent education systems (IESs), available for round-the-clock use, aiming to provide more personalized learning services to assist in automated teaching or learning support applications \cite{jiang2024intotheunknown,park2024empowering, chen2024dracademy, wang2023strategize,gao2025agent4edu,xu2024eduagent,soonwoo2024biped,qian2023user}. 

\begin{figure}[t]
    \centering
    \subfigure[Existing LLM-powered IESs.]{
        \includegraphics[width=\linewidth]{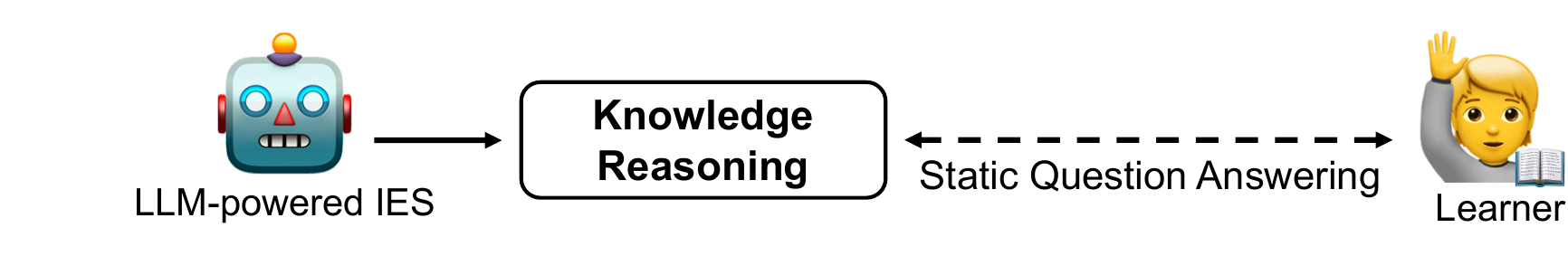}
        \label{fig:intro}
    }
    \vskip 0.1in
    \subfigure[Our proposed AgentTutor.]{
        \includegraphics[width=\linewidth]{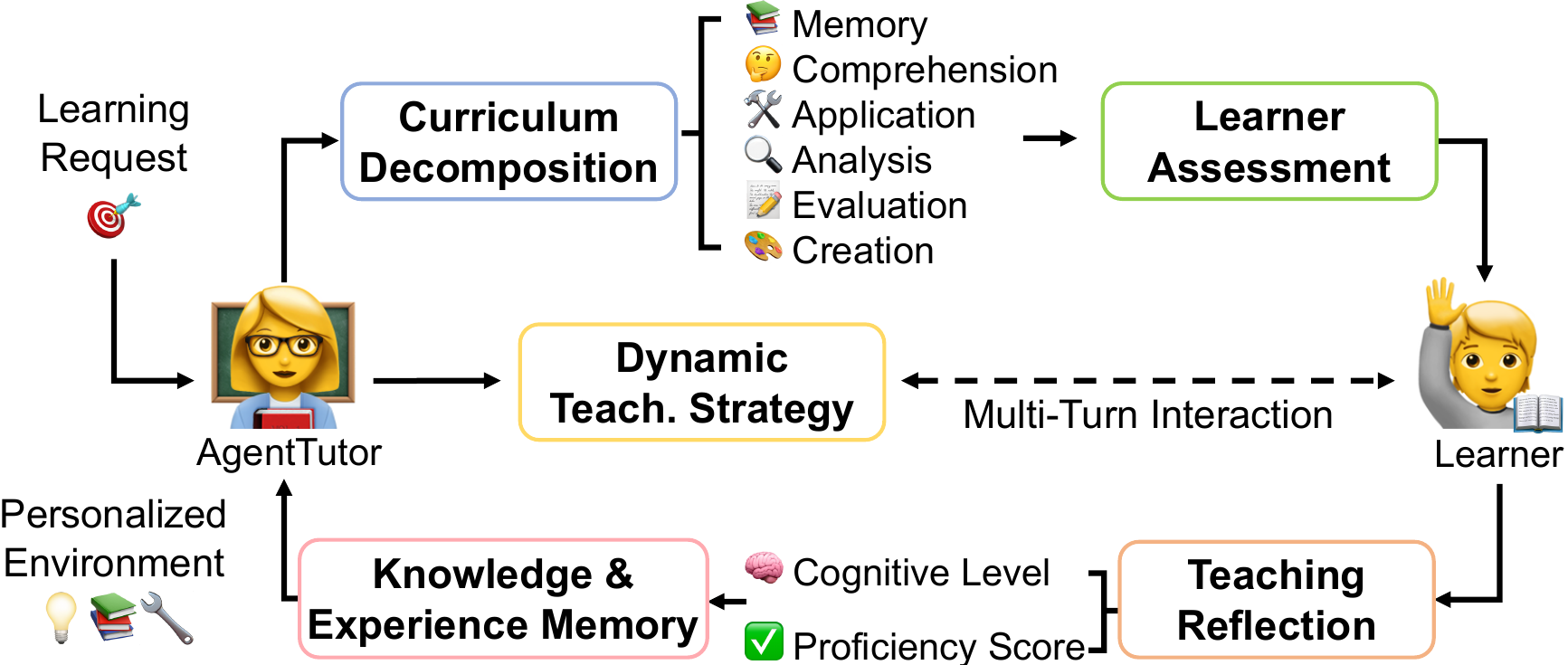}
        \label{fig:overview}
    }    \caption{A comparative analysis of existing systems and proposed AgentTutor. AgentTutor incorporates multiple LLM-powered generative agents along with a personalized environment to facilitate learning contexts.} \label{fig:intro_comparison}
    \vspace{-15pt}
\end{figure}

Personalized learning services focus on customized learning content, real-time feedback, and interactive experiences \cite{park2024empowering}. This adaptability is reflected in the ability to provide systematic guidance toward learners' learning curriculum, construct knowledge frameworks, and decompose complex problems based on learners' profile information. Tutors are required to continuously assess learners' cognitive levels and provide feedback, dynamically adjusting teaching strategies. Therefore, the personalized teaching process is a complex task involving sequential decision-making and multi-turn interactions.

Although existing LLM-powered IESs have achieved significant success in handling static single-turn question answering and knowledge reasoning tasks, these studies typically focus on answering specific, narrowly defined problems, as illustrated in Figure \ref{fig:intro}. Current systems, due to the absence of multi-turn interactive teaching, are unable to comprehensively assess learners' cognitive levels and learning progress, nor can they dynamically adjust teaching strategies based on real-time feedback. This limitation in multi-turn interaction hinders the effectiveness of these systems in more complex educational contexts, particularly in challenging scenarios like zero-shot inferences.

To address these issues, we propose AgentTutor, a multi-turn interactive intelligent education system designed for personalized learning, as shown in Figure \ref{fig:overview}. Inspired by Bloom's Taxonomy \cite{bloom2010taxonomy} and anthropological mechanisms \cite{wang2023user}, AgentTutor decomposes complex curriculum into sub-goals, continuously assesses learner progress, dynamically adjusts teaching strategies using tree search tools, and refines teaching through reflection on past experiences and knowledge. The system features an LLM-powered generative agent and an external learner-specific personalized learning profile environment. The generative agent provides dynamic teaching strategies based on the learner's cognitive level in a multi-turn process, comprising five modules: \textit{curriculum decomposition}, \textit{learner assessment}, \textit{dynamic strategy}, \textit{teaching reflection}, and \textit{knowledge \& experience memory}. 
The personalized learning environment enables learners to configure their learning curriculum, learning preferences, and multimodal study materials, allowing the generative agent to engage in direct, multi-turn interactive teaching. Our contributions are summarized as follows:

\begin{itemize}
    \item We developed AgentTutor, a multi-turn interactive intelligent education system. It applies multiple LLM-powered generative agents that can automate teaching strategies and provide dynamic, multi-turn teaching processes.
    
    \item AgentTutor is designed for dynamic, personalized learning scenarios, extending beyond single-turn static question answering. It integrates five core modules: \textit{curriculum decomposition}, \textit{learner assessment}, \textit{dynamic strategy}, \textit{teaching reflection}, and \textit{knowledge \& experience memory}, allowing for dynamic strategies, continuous assessment, and interactive experiences.

    \item We conducted extensive experiments on benchmark datasets and baselines. The system's effectiveness was evaluated via learners' improved performance, interactive teaching quality, human evaluations, and ablation studies, confirming the strong competitiveness of AgentTutor.
    
\end{itemize}

\section{Related Work}
\subsection{Intelligent Education Systems}
Intelligent Education Systems (IESs) offer personalized, real-time learning through computer-based technologies. Early rule-based systems like SCHOLAR \cite{carbonell1970ai} and ACT \cite{anderson1995cognitive} guided learners through predefined materials without human instructors, pioneering adaptive teaching. However, they couldn't dynamically adjust to learners' needs. With technological advancements, data-driven systems like knowledge tracing and emotional state recognition models \cite{corbett1994knowledge, d2008emotion} improved personalization by analyzing learner behavior. Yet, these systems still lacked deep understanding and high-level reasoning. The introduction of deep learning technologies further advanced IESs. Some researchers introduced a knowledge tracing model \cite{piech2015deep} by recurrent neural networks to model learners' learning levels, significantly improving prediction accuracy. Recently, generative AI models such as ChatGPT \cite{achiam2023gpt} and Gemini \cite{anil2023gemini} have shown great potential in personalized learning. However, they still face challenges in problem decomposition, multi-turn dynamic learning, and adjusting strategies based on ongoing learner progress, limiting their effectiveness in complex educational tasks \cite{baker2016stupid,huang2024enhancing,jurenka2024towards}.

\subsection{LLM-powered Agent Systems}
Many studies have utilized LLM-powered generative agents combined with domain-specific knowledge to design personalized applications in various scenarios. For instance, communications between users and customer service \cite{niu2024enhancing}, patient-doctor dialogues \cite{schmidgall2024agentclinic}, scriptwriting drama \cite{tu2024charactereval}, debates \cite{park2024predict}, and interactions between directors and actors \cite{han2024ibsen}. Recently, generative agent technologies have been applied to the field of education. For example, agents have been used to extract learners' cognitive factors to simulate the learning process \cite{xu2024eduagent,gao2025agent4edu,park2024empowering,www2025genmentor}, generate and summarize functional reports automatically \cite{jiang2024intotheunknown}, enhance the ability to generate high-quality standard questions based on educational theories \cite{chen2024dracademy}, and predict teachers' teaching strategies for data annotation \cite{wang2023strategize}. Although such progress in existing work has made, 
it often focuses on the implementation of fixed strategies and provides general teaching content, overlooking the dynamic changes in interaction and decision-making throughout the educational process. Therefore, our work focuses on the dynamic adjustment of teaching strategies based on the learner's cognitive level and learning progress.

\section{AgentTutor}

\begin{figure*}[ht]
    \centering\includegraphics[width=0.8\linewidth]{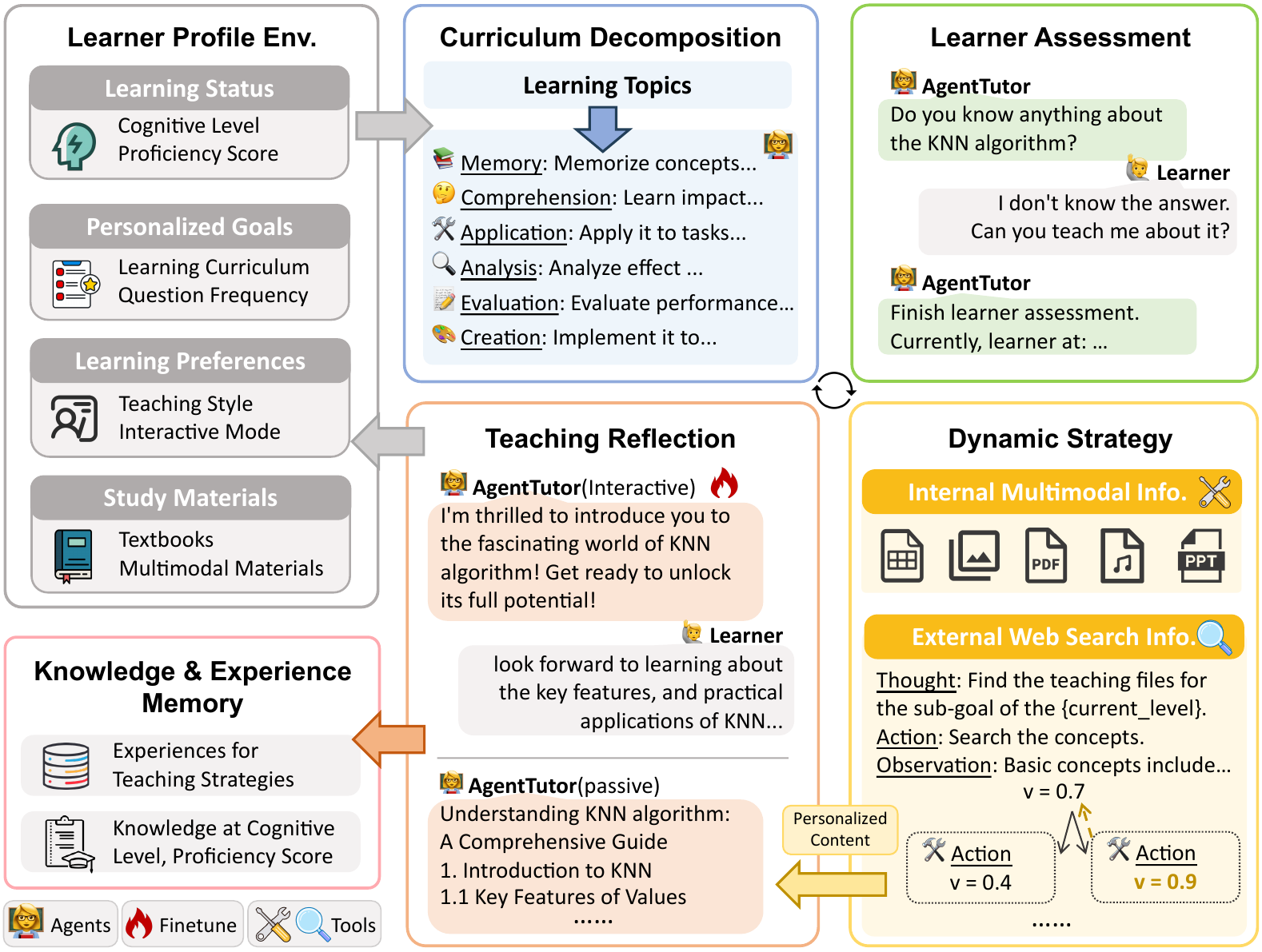}
     \caption{The diagram highlights the multi-turn interaction framework of the AgentTutor system, showing the collaboration among the five modules and personalized learning profile environment, forming a dynamic process that enhances teaching effectiveness. We take the KNN (K-Nearest Neighbors) algorithm as the curriculum example.}
    \label{fig:framework}
    \vspace{-10pt}
\end{figure*}

We propose AgentTutor to address multi-turn teaching interaction, featuring multiple LLM-powered generative agents and a personalized learning environment that configures curriculum goals, materials, and learning preferences based on the learner's profile. Operating as a professional tutor, the system integrates five core functional modules in a collaborative framework in Figure \ref{fig:framework} to form a dynamic multi-turn teaching loop: curriculum decomposition module decomposes the high-level curriculum into manageable sub-goals; learner assessment module continuously assesses the learner's cognitive level via questioning; dynamic strategy module leverages the LATS algorithm \cite{zhou2024lats} and multimodal resources to dynamically generate and optimize adaptive teaching strategies and content; teaching reflection module evaluates strategy effectiveness by generating practice tasks and feedback, using passive output or active conversational interaction; and knowledge \& experience memory module stores teaching experiences and learner archives as the system's long-term memory for future decision support.

\begin{figure*}[ht]
    \includegraphics[width=0.8\linewidth]{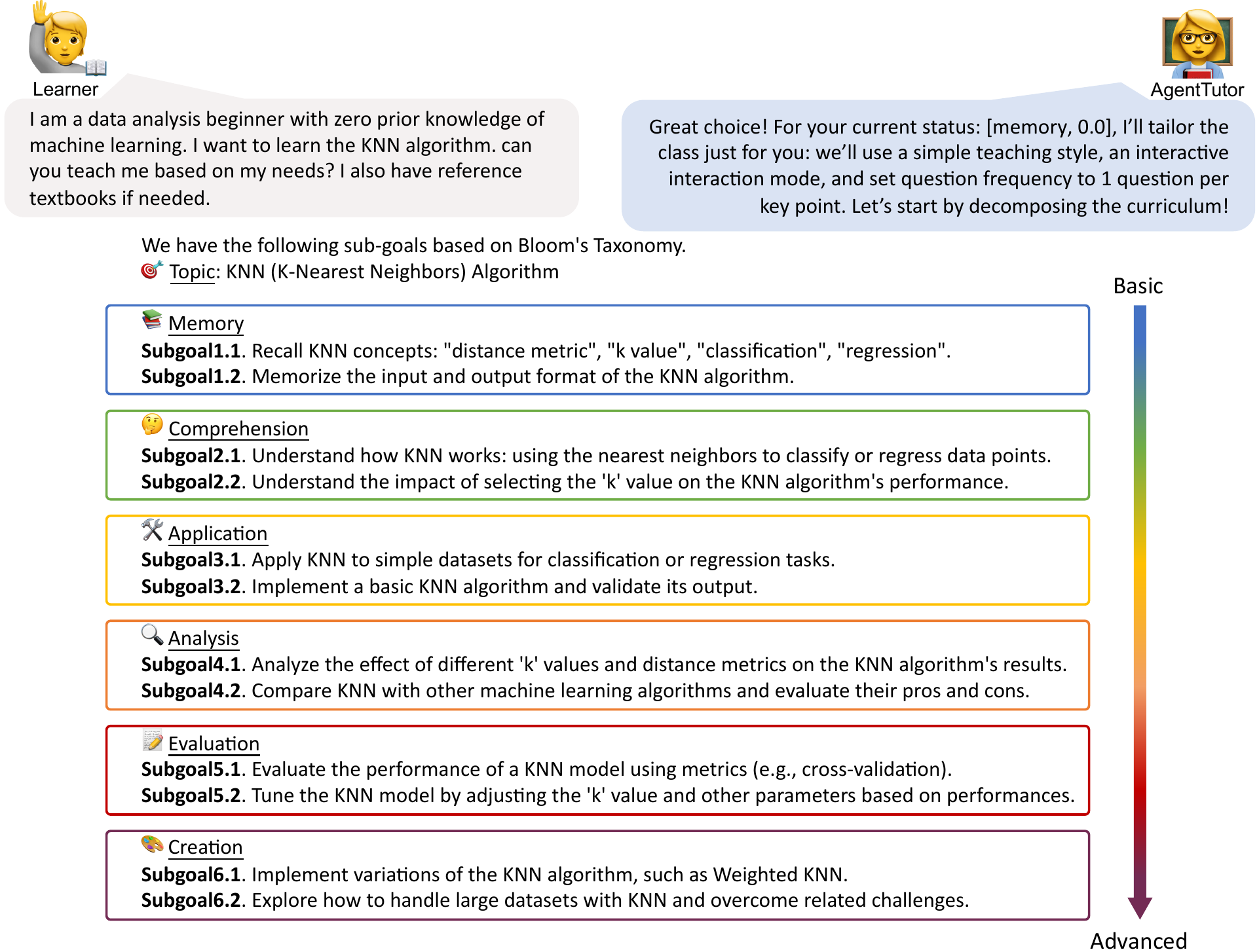}
    \centering
     \caption{Illustrating how the curriculum decomposition module decomposes a high-level curriculum, such as the KNN algorithm, into manageable sub-goals based on Bloom's Taxonomy. Leveraging educational theories, this module constructs a hierarchical tree that organizes these sub-goals into a structured learning pathway.}
    \label{fig:cdm_example}
    \vspace{-10pt}
\end{figure*}

\subsection{Problem Formulation}

Before introducing each module of AgentTutor, we first define the teaching strategy. Let $L$, $T$, and $G$ represent the cognitive level state space, teaching strategy space, and curriculum goal space, respectively. An educational process problem is formalized as a quintuple $P = (L, S, I, G, l_0)$, where $I: L \times S \rightarrow L$ denotes the state transition function and $l_0 \in L$ is the initial learning state. The objective is to determine an optimal teaching strategy $\pi: L \times G \rightarrow S$, for a given learning curriculum goal $g \in G$ and initial state $l_0$, the sequence of actions generated by executing strategy $\pi$ ultimately reaches as close as possible to the target state. Formally, we can express the problem as, $\pi^* = \operatorname{argmax}_{\pi} \mathbb{E}\left[ R(l_T, g) \mid l_0, \pi \right]$, where the reward function $R(l_T, g)$ measures the approximation degree of the final learning level state $l_T$ to the goal $g$ and $\mathbb{E}$ denotes the expectation.

\subsection{LLM-powered AgentTutor}

The generative agent in our system leverages foundational educational theories, the zero-shot capabilities of specialized generative agents, multimodal information integration, and tree-based search techniques to assess learners' cognitive levels and deliver dynamic, customized multi-turn teaching through the integration of five core modules.

\paragraph{Curriculum Decomposition Module (CDM).}  

The CDM serves as the foundation by decomposing a large curriculum, which includes high-level knowledge, into smaller, more manageable sub-goals. The primary objective of this design is to support systematic learning, enabling learners to progress from basic to advanced levels in a gradual and coherent manner, while also structuring the curriculum in a way that facilitates multiple turns of teaching and enhances the effectiveness of learner assessments and strategy searches. 

In accordance with established educational taxonomies \cite{bloom2010taxonomy}, the decomposition process is guided by six key domains: memory, comprehension, application, analysis, evaluation, and creation. Each of these domains serves as the basis for a specific goal, ensuring that the curriculum is comprehensive and covers various cognitive levels. Each high-level goal is further decomposed into a series of smaller sub-goals that align with six domains, as illustrated in Figure \ref{fig:cdm_example}. More details are provided in Appendix \ref{app:cdm}.

When a learner requests a learning curriculum, the initial process can be mathematically represented as follows. Let $G$ be the set of all learning goals, and $g_i \in G$ be a specific learning goal within the educational taxonomy. The decomposition function $D(g_i)$ can be defined as: $D(g_i) = \{b_1, b_2, ..., b_n\}$, where $b_j$ represents a sub-goal of $g_i$, and $n$ is the number of sub-goals. The decomposition function dynamically generates output based on LLM's semantic analysis. 

\paragraph{Learner Assessment Module (LAM).}  
The LAM iteratively evaluates the learner's cognitive level and learning progress. The purpose of this module is to align the sub-goals generated by the CDM, with the learner's current learning progress, ensuring that the subsequent teaching strategies are tailored to meet the learner’s abilities.

The process begins with an initial assessment, where LAM generates baseline questions to evaluate the learner’s foundational knowledge. This helps determine the learner's cognitive level and assign an appropriate proficiency score. In subsequent iterations, LAM continuously generates questions, aligned with the sub-goals and the learner’s evolving abilities. More details are provided in Appendix \ref{app:lam}.

The learner assessment function $A(b_j)$ is defined as $A(b_j) = (l_j, p_j)$, where $l_j$ is the Bloom's Taxonomy level (memory, comprehension, application, analysis, evaluation, creation), and $p_j \in [0, 1]$ is the proficiency score for the sub-goal $b_j$. The learner assessments form a directed acyclic graph $G = (V, E)$, where $V = \left\{(b_j, l_j, p_j) \mid b_j \in D(g_i)\right\}$ represents the sub-goals, their levels, and scores, and $E$ denotes the dependencies between sub-goals. The proficiency score $p_j$ is calculated as the average of the individual evaluation functions: $p_j = \frac{1}{m} \sum_{k=1}^m \beta_k \cdot f_k(b_j, r)$, where $r$ is the learner's response, $f_k$ are the evaluation metrics (e.g., accuracy, understanding, application), $\beta_k$ are the weights, and $m$ is the number of metrics.

\begin{algorithm}[t]\small
\caption{DSM Web Search Process}
\label{alg:simplified_dsm}
\begin{algorithmic}[1]
\REQUIRE Initial state $s$ in strategy space $S$, action generator $p_\theta$, rollouts number $K$, depth limit $D$, exploration weight $w$
\STATE Initialize strategy space $S$ with initial state $s$, value function $U$, and visit counter $N$
\FOR{$k = 1, \dots, K$}
    \FOR{$t = 1, \dots, D$}
        \IF{$s_t$ is not terminal}
            \FOR{$i = 1, \dots, n$}
                \STATE Sample action $a_{t,i} \sim p_\theta(s_t)$
                \STATE Simulate next $s_{t+1,i}$, evaluate $U_{t,i}$
                \STATE Update state-action value $U(s_t)$ based on feedback
            \ENDFOR
        \ENDIF
        \IF{$s_t$ is terminal}
            \STATE Reflect using reflect generator
        \ENDIF
        \STATE Calculate UCT score for current action: \\
        $
        UCT(s_t) = U(s_t) + w \cdot \sqrt{\frac{\ln N(s_t)}{N(s_{t+1})}}
        $
        \STATE Select best action: \\
        $
        a_t = \arg \max_{a \in A} \left[ \, UCT(s_t) \, \right]
        $
    \ENDFOR
    \STATE Backpropagate rewards: \\
    $U(s_t) \gets \frac{U(s_t)(N(s_t)-1) + r}{N(s_t)}$
\ENDFOR
\end{algorithmic}
\end{algorithm}

\paragraph{Dynamic Strategy Module (DSM).}  
The DSM comprises two components: first, the integration of internal study materials provided by the user via multimodal LLMs and tools like OCR, PDF readers, etc., and second, the search for external information relevant to the current cognitive level, which is dynamically generated by iteratively executing the language agent tree search algorithm. The final output is a ready-to-use lesson plan or plug-and-play lesson content compiled from all relevant information. The core idea of DSM is to integrate appropriate resources and adapt strategies based on the learner's cognitive level and learning sub-goals, ensuring that strategies are continuously refined.

The DSM treats the external information generation process process as a tree-based exploration problem, where each node represents a potential teaching decision, and edges denote transitions between actions. LATS, an advanced web search technique, combines generation, reasoning, reflection, and optimization to create teaching strategies that cater to each learner’s learning level \cite{zhou2024lats}. The key operations of DSM shown in Algo. \ref{alg:simplified_dsm}, involve selecting promising strategies using the Upper Confidence Bound for Trees (UCT) score, expanding them into multiple candidates, simulating their effectiveness based on the learner's learning level, evaluating them with internal and external feedback, refining strategies through backpropagation, and integrating external tools. More details are in Appendix \ref{app:dsm}.

The mathematical representation is as follows: Let $M$ represent the search tree, where $s \in S$ is a teaching strategy node and $a \in A$ is a search action. The UCT score for selecting node $s$ is computed as: $UCT(s) = U(s) + w \sqrt{\frac{\ln N(p)}{N(s)}}$, where $U(s)$ is the estimated value of the node, $N(s)$ is the visit count of the node, $N(p)$ is the visit count of the parent node, and $w$ is the exploration weight that balances exploration of new strategies and exploitation of known efficient strategy. The value function $U(s)$ is composed of the score derived from the LLM's output, and the self-consistency score, ensuring logical coherence and accuracy.

\paragraph{Teaching Reflection Module (TRM).} The TRM aims to provide either conversational teaching or direct learning files based on different interaction modes. After this, it generates practice tasks based on the question frequency, assesses learner performance, and provides tailored feedback. The goal is to provide personalized instruction and ensure continuous refinement of teaching methods based on the learner's evolving progress, thereby promoting ongoing improvement in their learning process. By tailoring the instructional mode and feedback to each learner's performance, the module supports personalized learning that adapts to individual growth.

The conversational model here trained using offline reinforcement learning, specifically Direct Preference Optimization (DPO) on tutoring datasets~\cite{macina2023mathdial,scarlatos2025trainingllmbasedtutorsimprove}, aims to maximize the likelihood of eliciting correct learner responses while adhering to pedagogical principles, demonstrating improvement in accuracy, progress tracking, error identification, strategic hinting, and encouraging. The interaction mode can be dialogue or passive reading based on the corresponding teaching content. Additionally, teaching strategies reflection is primarily evaluated by providing targeted tasks. For the programming curriculum, for high-performing learners, this module generates more complex tasks, while lower-performing learners are given tasks that provide foundational guidance. Upon task completion, the TRM evaluates the learner’s performance using standard grading criteria, which assess the correctness, efficiency, and maintainability of the task \cite{tong2024codejudge}. The evaluation criteria of AgentTutors are orthogonal, meaning they can be customized or integrated with any other metrics based on the learner’s material requirements. More details are provided in Appendix \ref{app:trm}.

\paragraph{Knowledge \& Experience Memory Module (KEMM).}  
The KEMM serves as the AgentTutor’s long-term memory, storing and retrieving learner knowledge archives and teaching experiences. Inspired by anthropological memory mechanisms \cite{brown2020language, wang2023user}, KEMM ensures that the system retains valuable past interactions to inform future teaching decisions. This design facilitates continuous learning, enabling the system to adapt and evolve based on accumulated knowledge.

To efficiently manage this knowledge, KEMM utilizes a vector database with a semantic indexing structure for quick information retrieval and updates. This process enhances the DSM, providing the prior learning data for improving teaching strategies. When AgentTutor generates teaching strategies in subsequent interactions, it can reference the long-term memory of the learner's knowledge and the corresponding teaching experiences, enabling rapid retrieval of relevant strategies. 

\subsection{Learner Profile Environment.}  
The personalized profile environment enables learners to configure various curricula, such as mathematical reasoning, coding problems, and language learning, based on their current cognitive level and personalized learning goals. It also allows for the provision of personal textbooks or other multimodal study materials and preferred instructional modes. This design is intended to provide tailored and adaptive learning contexts based on the learner's needs. For illustration, we selected coding problems to demonstrate how the system provides systematic guidance and feedback around a broader educational theme in the personalized context as shown in Figure \ref{fig:framework}. More details are provided in Appendix \ref{app:case_study}.

\section{Experiments}
We conduct experiments concerning the following research questions:
\begin{enumerate}[leftmargin=25pt]
    \item What is the overall teaching effectiveness after tutor and learner interactions within IES?
    \item How does AgentTutor perform in terms of interactive teaching quality?
    \item How do human experts evaluate the teaching experience provided by AgentTutor?
    \item What is the contribution of each module?
\end{enumerate}

\paragraph{Datasets.} We utilized two publicly available datasets based on programming courses. HumanEval \cite{chen2021humaneval} consists of $164$ programming tasks primarily used for evaluating coding abilities. MBPP \cite{austin2021mbpp} contains $974$ programming tasks and is suitable for assessing entry-level programming skills. We chose programming curricula for our experiments due to their challenging nature, the ability to cover diverse scenarios and learners, and the availability of multiple baselines for comprehensive evaluation.

\paragraph{Baselines.} We selected five baselines to evaluate the performance of reasoning methods in IESs, including Chain of Thought (CoT) \cite{wei2022chain}, ReAct \cite{yao2022react}, Tree of Thought (ToT) \cite{yao2024tree}, Reasoning with Action Planning (RAP) \cite{hao2023reasoning}, and Reflexion \cite{shinn2023reflexion}. 

\setlength{\cmidrulewidth}{0.1mm} 
\renewcommand{\arraystretch}{1} 
\begin{table}[t]
    \centering
    \begin{tabular}{l c c }
    \toprule
        \multicolumn{1}{c}{} & \multicolumn{2}{c}{\textbf{HumanEval}} \\
        \cmidrule(lr){2-3}
        \textbf{Method} & \textbf{Model} & \textbf{Pass@1} \\ \hline
        CoT  & GPT-3.5-turbo & 46.9 \\
         ReAct & GPT-3.5-turbo & 56.9 \\
         ToT & GPT-3.5-turbo & 54.4  \\
         RAP & GPT-3.5-turbo & 63.1 \\
         Reflexion & GPT-3.5-turbo & 68.1 \\
         AgentTutor & GPT-3.5-turbo & 92.7 \\ \hline
         Zero-shot & GPT-4& 80.1  \\
         Reflexion & GPT-4& 91.0  \\
         AgentTutor & GPT-4& \textbf{96.9}  \\ \bottomrule
    \end{tabular}
    \caption{Comparison of Pass@1 accuracy on the HumanEval benchmark for GPT-3.5-turbo, GPT-4, and AgentTutor. AgentTutor demonstrates the highest performance among all evaluated methods.}
    \label{tab:humaneval}
    \vspace{-10pt}
\end{table}

\paragraph{Experimental Setting.} We designed the system using the open-source LangChain framework \cite{langchain} and ChromaDB database \cite{chromadb} as the knowledge base. To enrich the teaching system with knowledge, we primarily employed the OpenAI API for conversation and Qwen2.5-VL-7B-Instruct as the multimodal usage. Considering cost constraints, for tasks requiring reasoning, we used high-performance models like GPT-4 (annotated in subsequent experiments); for other conversational interactions, we used GPT-3.5-turbo by default, setting the temperature to $0.7$. For the LATS algorithm, we controlled the number of candidate answers to $3$, the maximum tree search depth to $3$, and the candidate answer quality threshold to $7$, while utilizing the Tavily tool \cite{tavily} for web search, and other local tools for OCR and PDF reading. Additionally, we employed GPT-3.5-turbo for the learner agent, which exhibits three behavioral states (confusion, learning, response) following the setting by \citet{macina2023mathdial}, and detailed descriptions are provided in Appendix \ref{app:learner}. The learner's cognitive level is initialized to the basic level (memory), with a proficiency score of $0$. To validate our system's capabilities, we rely solely on web searches for teaching strategies without additional materials. For each theme in the benchmark datasets, AgentTutor engages with each learner for $10$ turns. LLM settings for baseline methods were kept consistent with ours. Regarding personalized settings, we uniformly set question frequency to high, teaching style to detailed, and interactive mode to passive. For training the conversational model, we referenced the parameter settings in \citet{scarlatos2025trainingllmbasedtutorsimprove}, using Llama-3.1-8B-instruct for LoRA training with rank $64$, scaling factor $32$, dropout rate of $0.05$, the AdamW optimizer, batch size $64$, and a decay weight of $0.01$. All experiments were conducted on the NVIDIA A800 GPUs.

\subsection{Multi-Turn Teaching Effectiveness}

To thoroughly assess the effectiveness of the AgentTutor system, we primarily focus on the improvement in learner performance following interactions with the system. Our evaluation is based on the Pass@1 metric, defined as the percentage of generated code that passes all test cases on the first attempt. This metric serves as a robust indicator of both the correctness and efficiency of the solutions produced by the learner agent after engaging with the system. All results are presented as averages.

Table \ref{tab:humaneval} and Table \ref{tab:mbpp} show that the learner achieves the highest improvements in Pass@1 on the HumanEval and MBPP datasets. The results demonstrate that the enhanced dynamic adjustment of teaching strategies and multi-turn interactions significantly contribute to better learner performance. Unlike other baseline methods, our system, leveraging the LATS algorithm, utilizes a tree-based search combined with reasoning, self-reflection, and memory to identify high-quality teaching strategies and materials that align with the learner's current cognitive level. These dynamic strategies allow for more targeted guidance, while the multi-turn interaction design enables learners to progress from basic to more advanced concepts in a comprehensive manner. 

\subsection{Interactive Teaching Quality}

\begin{table}[t]
    \centering
    \begin{tabular}{l c c }
    \toprule
        \multicolumn{1}{c}{} & \multicolumn{2}{c}{\textbf{MBPP}} \\
        \cmidrule(lr){2-3}
        \textbf{Method} & \textbf{Model} & \textbf{Pass@1}  \\ \hline
         CoT &  GPT-3.5-turbo & 54.9 \\
         ReAct &  GPT-3.5-turbo & 67.0 \\
         ToT &  GPT-3.5-turbo & 65.8 \\
         RAP &  GPT-3.5-turbo & 71.4 \\
         Reflexion &  GPT-3.5-turbo & 70.0 \\
         AgentTutor &  GPT-3.5-turbo & \textbf{89.4} \\
         \bottomrule
    \end{tabular}
    \caption{Comparison of Pass@1 accuracy on the MBPP benchmark for baselines. AgentTutor demonstrates the highest performance among all evaluated methods.}
    \label{tab:mbpp}
    \vspace{-10pt}
\end{table}

We now evaluate the quality of the interactive teaching in a multi-turn process. We adapt the automated metrics to measure the extent to which the teaching interaction adheres to effective pedagogical practices \cite{jurenka2024towards}, such as appropriate responses, answer feedback, question formulation (i.e., Conversational Adaptability, Feedback Quality, and Question Difficulty). We utilize Prometheus 2 \cite{kim2024prometheus2}, a 7B evaluator LLM, to score the interaction process based on a 5-point rubric. The experimental details are provided in the Appendix \ref{app:automated_metric}. 

\begin{figure}[t]
    \centering
    \includegraphics[width=0.8\linewidth]{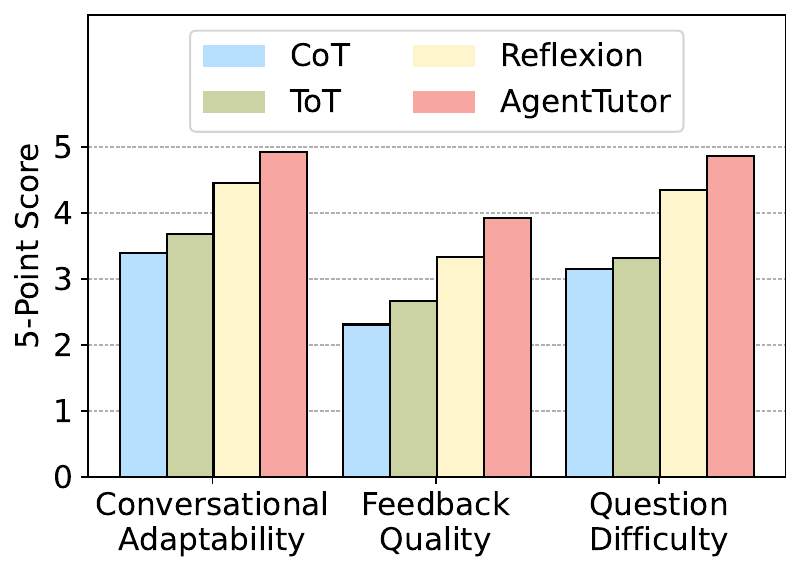}
    \caption{Comparison of interactive teaching quality across multi-turn interactions by conversational adaptability, feedback quality, and question difficulty.}
    \label{fig:quality}
     \vspace{-10pt}
\end{figure}

Experimental results in Figure \ref{fig:quality} show that AgentTutor outperforms other methods across all metrics. CoT, ToT, and Reflexion are primarily designed for single-turn interactions, they do not adapt the learner’s previous responses, leading to less relevant teaching strategies, lower quality feedback, and less appropriately formulated questions to effectively guide the learner's progress. In contrast, AgentTutor’s multi-turn interaction design, which incorporates curriculum decomposition, learner assessment, and dynamic teaching strategies, ensures better alignment with the learner’s cognitive level and a more effective conversation progression through Bloom’s Taxonomy, thereby significantly enhancing interactive teaching quality.

\subsection{Human Evaluation}

\begin{figure}[t]
    \centering
    \includegraphics[width=0.7\linewidth]{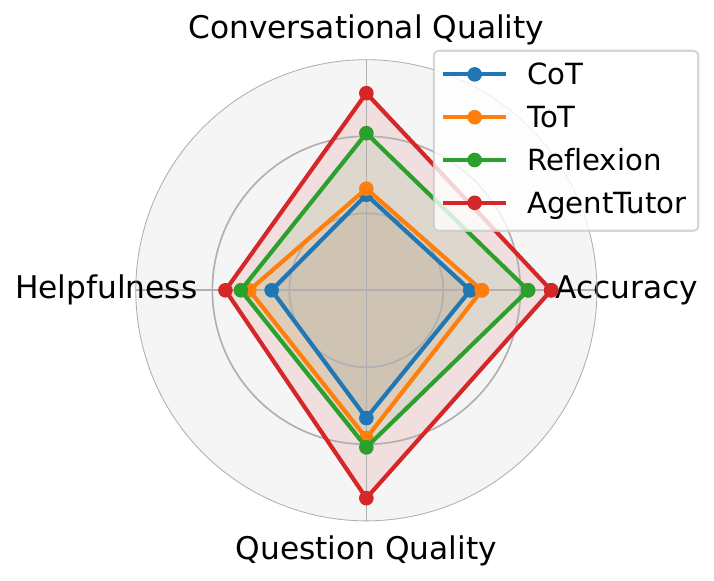}
    \caption{Human evaluation of various teaching methods based on accuracy, conversational quality, helpfulness, and question set quality. AgentTutor demonstrates the highest performance across all criteria.}
    \label{fig:human}
\end{figure}

We further conducted human evaluations on the methods in Table \ref{fig:human} based on four criteria introduced by \citet{jurenka2024towards}: (1) Accuracy: including assessment and feedback accuracy; (2) Conversational Quality: encompassing teaching engagement, response length, and personalized learning context usage; (3) Helpfulness and Relevance: evaluating the relevance and helpfulness of the tutor's feedback; and (4) Question Set Quality: assessing how well the question set is formulated relative to the current learner cognitive level. Three teaching experts voluntarily participated in the study and consented to provide in-depth feedback on the pedagogical value. They rated each item on a 3-point scale based on the criteria outlined above. Our findings indicate that the multi-turn interactive system achieves favorable scores in human evaluations. The result of AgentTutor excels because of its multi-turn interaction design, enabling better engagement, more relevant feedback, and higher quality questions that align with the learner’s level.

\subsection{Ablation Study}  

We evaluate the impact of each module, as shown in Table \ref{tab:ablation}. All modules significantly contribute to the reasoning process and the generation of dynamic teaching strategies. DSM and CDM are particularly critical. Removing DSM causes a 19.8\% performance drop, highlighting its role in finding better answers using web search tools. CDM, which breaks down complex problems, results in an 8.6\% drop when omitted, emphasizing its importance in structuring the teaching process. Other modules also contribute, reinforcing the system’s effectiveness in achieving strong teaching outcomes.

\begin{table}[ht]
    \centering
    \begin{tabular}{@{}lcc@{}}
    \toprule
    \textbf{Method} & \textbf{Model} & \textbf{Pass@1} \\ \midrule
    AgentTutor &GPT-3.5-turbo& \textbf{92.7}  \\
    w/o DSM &GPT-3.5-turbo& 72.9 (-19.8) \\
    w/o CDM &GPT-3.5-turbo& 84.1 (-8.6)\\
    w/o KEMM &GPT-3.5-turbo& 85.3 (-7.4) \\
    w/o LAM &GPT-3.5-turbo& 85.7 (-7.0) \\
    w/o TRM &GPT-3.5-turbo& 86.2 (-6.5) \\ \bottomrule
    \end{tabular}
    \caption{Ablation Study Results for GPT-3.5-turbo Pass@1 accuracy on HumanEval.}
    \label{tab:ablation}
    \vspace{-10pt}
\end{table} 

\section{Challenges}
While AgentTutor shows progress in IESs, we acknowledge some challenges limiting its potential. Firstly, evaluation metrics are insufficient: The lack of quality multi-turn interaction datasets and the inadequacy of domain-independent metrics like BLEU impede robust pedagogical assessment. Our current evaluation relies on subjective or costly LLM evaluators and human experts. Second, personalization scope is Narrow: The system primarily focuses on the learner's cognitive level and progress score, with case studies centered on the specific domain. Future iterations must broaden the scope to include more factors across diverse subjects. Third, real-world learner validation is limited: Due to resource constraints on large-scale trials, our reliance on simulated learner affects the generality of our findings. Fourth, multimodal generation requires development: Although multimodal information can be extracted and integrated, the challenge remains in developing effective methods for generating high-quality, pedagogical multimodal lesson plans from this information.

\section{Conclusion}
In this paper, we introduced AgentTutor, a multi-turn interactive education system powered by LLMs. Unlike single-turn static question-answering systems, AgentTutor provides continuous assessment, adapts dynamic strategies, and engages in interactive experiences, offering personalized learning based on the learner profile environment. Our experiments show that AgentTutor outperforms existing systems across several benchmarks, particularly in the learner's performance improvement and multi-turn interaction quality. Human evaluations confirm its engaging, relevant, and effective teaching, highlighting the potential of multi-turn interactive systems in intelligent education.

\section*{Acknowledgment}
This work was supported in part by National Key R\&D Program of China No. 2024YFB2705300, Research Grants from State Key Laboratory of Submarine Geoscience, NSFC Grant 62232011, 62402315, and the Shanghai Science and Technology Innovation Action Plan Grant 24BC3201200.

\bibliography{LaTeX/aaai2026}

\newpage
\onecolumn
\appendix

\section{Appendix} \label{sec:app}
\subsection{CDM Prompt Design} \label{app:cdm}
In AgentTutor system, the CDM uses a structured prompt to break down a topic based on Bloom's Taxonomy into manageable learning sub-goals. The prompt takes the learner's question as input, and the detailed prompt description is shown in Table \ref{tab:cdm_input}. Then, CDM outputs a JSON format sub-goals organized by cognitive levels, and we take KNN algorithm as an example shown in Figure \ref{fig:cdm_example} and Table \ref{tab:cdm_output}.

\setlength{\cmidrulewidth}{0.1mm} 
\renewcommand{\arraystretch}{1} 
\begin{table}[ht]
\centering 
\begin{tabular}{p{16cm}}
\toprule
\textbf{Instruction:}
You are an intelligent teaching assistant who follows Bloom's Taxonomy-based Intelligent Decomposition methodology to help learners break down a topic into manageable learning sub-goals. When the learner wants to learn the $\{curriculum\}$, you will break it down into the following sub-goals based on cognitive levels. \\
\textbf{Learner's Input:} $\{curriculum\}$ \\
\textbf{Memory Level:}
Identify and list the basic concepts, definitions, and key terms the learner needs to memorize. These are the foundational facts necessary to understand the topic. \\
\textbf{Comprehension Level:}
Help the learner understand the core concepts of the topic. This involves explaining how things work, providing examples, and making sure the learner can explain it in their own words.\\
\textbf{Application Level:}
Guide the learner through using the concepts they have learned by solving real-world problems or exercises. Encourage hands-on practice and implementation of the concepts.\\
\textbf{Analysis Level:}
Encourage the learner to analyze the topic by comparing, contrasting, and evaluating the different approaches or aspects. Help them identify strengths, weaknesses, and potential improvements.\\
\textbf{Evaluation Level:}
Help the learner evaluate the performance or effectiveness of a method, model, or approach. This can include reviewing evaluation metrics, trade-offs, or optimal parameters.\\
\textbf{Creation Level:}
Encourage the learner to apply their knowledge creatively. Help them design new experiments, algorithms, or ideas that expand on the basic knowledge and application. \\ \bottomrule
\end{tabular} 
\caption{CDM prompt description, detailing the six levels based on Bloom's Taxonomy.} \label{tab:cdm_input}
\end{table}

\subsection{LAM Prompt Design}\label{app:lam}

The LAM evaluates a learner's answer based on Bloom's Taxonomy, considering their current level and the sub-goals associated with the curriculum topic. The process involves breaking down the assessment into different cognitive levels and assigning scores based on the learner's understanding. The system outputs a dictionary with two fields: the level that best corresponds to the learner's answer and the decimal score representing the learner's mastery of that level shown in Table \ref{tab:lam_prompt}. 

\setlength{\cmidrulewidth}{0.1mm} 
\renewcommand{\arraystretch}{1} 
\begin{table}[htbp]
\centering 
\begin{tabular}{p{16cm}}
\toprule
\textbf{Instruction:} \\
You are an intelligent teaching assistant tasked with evaluating a learner's answer based on Bloom's Taxonomy. 

The $\{question\}$ is based on the $\{current\_level\}$ and $\{current\_score\}$. Given the learner's answer $\{learner\_answer\}$, evaluate the $\{topic\}$ in ascending order of complexity and assign a mastery score from 0.0 to 1.0 for each level. \\
1. \textbf{Memory:} \\
- Does the learner's response reflect knowledge or recall? \\
- Sub-goals: $\{memory\}$ \\
2. \textbf{Comprehension:} \\
- Does it show an understanding of the topic? \\
- Sub-goals: $\{comprehension\}$ \\
3. \textbf{Application:} \\
- Has the learner applied the concept or used it in some practical context? \\
- Sub-goals: $\{application\}$ \\
4. \textbf{Analysis:} \\
- Has the learner analyzed or critically evaluated the topic? \\
- Sub-goals: $\{analysis\}$ \\
5. \textbf{Evaluation:} \\
- Has the learner evaluated the topic or made judgments based on criteria? \\
- Sub-goals: $\{evaluation\}$ \\
6. \textbf{Creation:} \\
- Does the learner demonstrate original thought or the ability to create new ideas, algorithms, or approaches to solve the problem? \\
- Sub-goals: $\{creation\}$ \\ \bottomrule
\end{tabular} 
\caption{LAM prompt description based evaluation of learner's answer.} \label{tab:lam_prompt}
\end{table}

\begin{algorithm*}[b]
\caption{LATS ($s$, $p_\varphi$, $p_\varphi$, $p_{ref}$, $n$, $D$, $K$, $c$, $w$, $a$, $\lambda$)}
\label{alg:lats}
\begin{algorithmic}[1]
\REQUIRE Initial state $s$ in teaching strategies space, action generator $p_\theta$, value function $p_\varphi$, reflection generator $p_{ref}$, number of generated actions $n$, depth limit $D$, number of rollouts $K$, context $c$, exploration weight $w$, and value function weight $\lambda$
\STATE Initialize action space $A$, observation space $O$
\STATE Initialize the state-action value function $p_\varphi : S \times A \mapsto \mathbb{R}$ and visit counter $N : S \mapsto \mathbb{N}$ to one
\FOR{$k = 0, \dots, K-1$}
    \FOR{$t = 0, \dots, D-1$}
        \IF{$s_t$ is not terminal}
            \FOR{$i = 1, \dots, n$}
                \STATE Sample $a_{t,i} \sim p_\theta(s_t)$ \hfill \textcolor{gray}{Expansion \& Simulation}
                \STATE Get $o_{t,i}$ from environment, $s_{t+1,i} \gets (c_i, o_{t,i}, a_{t,i})$, $c_{t+1,i} \gets (o_{t,i}, a_{t,i})$
                \STATE Evaluate $U_{t,i} \sim \lambda \cdot p_\varphi(s_{t,i}) + (1 - \lambda) \cdot SC(s_{t,i})$ \hfill \textcolor{gray}{Evaluation}
                \STATE $U(s_t) \gets U_{t,i}$
                \STATE Add $s_{t,i}$ to children
            \ENDFOR
        \ENDIF
        \IF{$s_t$ is terminal}
            \STATE Get $r$ from environment \hfill \textcolor{gray}{Reflection}
            \IF{$r$ is not success}
                \STATE reflection $\gets p_{ref}(c_t)$
                \STATE $c_t \gets c_t -$ reflection
            \ENDIF
        \ENDIF
        \STATE $a_t \gets \arg \max_{a \in A} \left[ U(s_t) + w \sqrt{\frac{\ln N(s_t)}{N(s_{t+1})}} \right]$ \hfill \textcolor{gray}{Selection}
        \STATE Get corresponding $o_t$ from memory, $s_{t+1} \gets (c_t, o_t, a_t)$, $c_{t+1} \gets (o_t, a_t)$
        \IF{$a_t$ is an output action}
            \STATE \textbf{break}
        \ENDIF
    \ENDFOR
    \STATE $T \gets$ the actual number of steps \hfill \textcolor{gray}{Backpropagation}
    \FOR{$t = T - 1, \dots, 0$}
        \STATE $U(s_t) \gets \frac{U(s_t)(N(s_t)-1)+r}{N(s_t)}$
    \ENDFOR
\ENDFOR
\end{algorithmic}
\end{algorithm*}

\subsection{DSM Algorithm Design}\label{app:dsm}

The DSM is designed to generate, evaluate, and refine teaching strategies. The first step involves summarizing the provided learning materials using a multimodal LLM configured with tools (e.g., OCR, PDF readers) to convert existing information into corresponding text. The second step uses the LATS algorithm for external web searches, simulating possible teaching actions and using feedback to improve AgentTutor's decision-making. Finally, all collected information is compiled into the necessary lesson plan. The complete search pseudocode, shown in Algorithm \ref{alg:lats}, defines the essential components, like action generator, value function, reflection generator, and backpropagation.  

The key operations of DSM include:  
(1) Selection: At each step, the system uses the Upper Confidence Bound for Trees algorithm to select the most promising node for expansion. This step balances the trade-off between discovering new strategies and refining known high-value strategies, thereby optimizing the overall teaching strategies.
(2) Expansion and Simulation: The selected node is expanded by generating multiple candidate teaching strategies (e.g., generating five candidates per iteration). These strategies are then simulated to predict their effectiveness, taking into account the learner's current learning level, goals, and previous feedback.  
(3) Reflection and Evaluation: Each candidate strategy is evaluated based on the simulation outcomes, followed by reflection. This module combines internal scoring (e.g., learner level) and external feedback (e.g., knowledge retrieved from the web or knowledge bases) to assess the alignment of the strategy with the learner’s needs and the sub-goal requirements. Specifically, the value function $U(s)$ is composed of two parts: $U(s) = \lambda \cdot p_\varphi(s) + (1 - \lambda) \cdot SC(s)$, where $p_\varphi(s)$ represents the score derived from the LLM's output, $SC(s)$ represents the self-consistency score, and $\lambda$ determines their relative influence.  
(4) Backpropagation: The evaluation results are propagated back through the search tree to update the value estimates of parent nodes. This process helps refine the trajectory for subsequent iterations by rewarding effective strategies and penalizing suboptimal ones.  
(5) Integration with External Knowledge: The DSM incorporates relevant external knowledge sources (e.g., other retrieved study materials) into the search process. These resources ensure the personalization and accuracy of the teaching plan by providing contextually relevant and up-to-date information.

In all experiments, we set the number of sampled nodes to $n = 3$ and the exploration weight to $w = 1$. We use a self-consistency weight of $\lambda= 0.8$ for programming experiments. 

\setlength{\cmidrulewidth}{0.1mm} 
\renewcommand{\arraystretch}{1} 
\begin{table*}[t]
\centering 
\begin{tabular}{p{2.5cm} l}
\toprule
\textbf{Criteria} & \textbf{Description} \\
\hline
\textbf{Functionality} & 
\begin{tabular}[c]{p{13cm}} 
- Does the code correctly implement the functionality as described in the question? \\
- Does it meet the requirements specified in the task description?
\end{tabular} \\
\hline
\textbf{Code Quality} & 
\begin{tabular}[c]{p{13cm}} 
- Does the code follow best practices, such as proper naming conventions, comments, and formatting? \\
- Is the code clean, readable, and well-structured?
\end{tabular} \\
\hline
\textbf{Performance} & 
\begin{tabular}[c]{p{13cm}} 
- Does the code implement an efficient algorithm in terms of time and space complexity? \\
- Are there any optimization opportunities in terms of algorithmic efficiency?
\end{tabular} \\
\hline
\textbf{Maintainability} & 
\begin{tabular}[c]{p{13cm}} 
- Is the code modular, easy to understand, and extendable? \\
- Can future modifications or additions be made with minimal effort or issues?
\end{tabular} \\
\hline
\textbf{Overall} & 
\begin{tabular}[c]{p{13cm}} 
- Negligible: The code has severe issues, such as missing imports, logical errors, or major inefficiencies. \\
- Small: The code has small issues, such as missing edge case handling or inefficient approaches. \\
- Major: The code has major issues like logical errors, missing features, or poor performance. \\
- Fatal: The code does not implement the required functionality at all or contains fatal errors.
\end{tabular} \\
\bottomrule
\end{tabular}
\caption{Evaluation metrics for code-based tasks in TRM. We take the HumanEval programming task as an illustrative example.}
\label{tab:code_evaluation_metrics}
\end{table*}

\setlength{\cmidrulewidth}{0.1mm} 
\renewcommand{\arraystretch}{1} 
\begin{table}[ht]
\centering 
\begin{tabular}{p{16cm}}
\toprule
\textbf{Instruction:} \\
Generate a task for a learner based on Bloom’s Taxonomy. The learner’s current level is $\{current\_level\}$. Given the sub-goals for this level, generate a task that aligns with the learner’s level and helps progress their understanding. Ensure the task is challenging but achievable based on their prior performance. \\

Evaluate the learner's $\{learner\_answer\}$ based on the sub-goals for each level. For each level, assign a mastery score between 0.0 and 1.0, based on the given evaluation criteria like programming domain (functionality, code quality, performance, and maintainability). Provide a score and a remark explaining your evaluation. \\

Generate a follow-up question for a learner based on the sub-goals. The learner’s current level is $\{current\_level'\}$. \\ \bottomrule
\end{tabular} 
\caption{TRM prompt description, including task generation, task evaluation, and question generation.} \label{tab:trm_input}
\end{table}

\subsection{TRM Prompt Design}\label{app:trm}

The TRM first integrates the teaching content collected by the preceding DSM. It then conducts instruction based on the learner's interactive mode (active or passive). Next, based on the learner's cognitive level and progress, it generates practice tasks, evaluates the learner's answer, and generates new questions for the next turn. The TRM consists of four core components in Table \ref{tab:trm_input}: (1) Interactive Teaching: In interactive mode, AgentTutor initiates a fine-tuned conversational model for dialogic instruction based on the content, while in passive mode, it directly displays the teaching files for reading based on the content. (2) Task Generation: Dynamically generates tasks based on the learner’s current knowledge and learning goals. (3) Task Evaluation: Assesses the learner’s responses based on Bloom's Taxonomy, assigns scores while considering prior performance, and provides feedback. (4) Question Generation: Creates follow-up questions or new tasks that are contextually appropriate to advance the learner’s understanding.

\setlength{\cmidrulewidth}{0.1mm} 
\renewcommand{\arraystretch}{1} 
\begin{table*}[t]
    \centering
    \begin{tabular}{p{3cm}p{13cm}}
        \toprule
        \textbf{Mode} & \textbf{Prompt Design} \\
        \midrule
        \textbf{Confusion} & 
        I don't know the answer. Can you teach me about it? \\
        \midrule
        \textbf{Learning} & 
        (No specific prompt is explicitly defined in this mode. Its main function is to add new knowledge to the learner's knowledge base, such as adding the content input by the tutor and documents loaded from URLs provided by the tutor.) \\
        \midrule
        \textbf{Response} & 
        You are a curious learner who only has knowledge stored in your knowledge base, which is like your brain.
        
        You have retrieved the following information from your brain KNOWLEDGE: $\{context\}$     

        Now, think carefully about the question asked: $\{question\}$. Use the information from your brain (the knowledge base) to guide your thinking.       

        If you know the answer based on what you remember, respond directly. 
        
        If you're not sure, try reasoning only based on KNOWLEDGE.
        
        If your KNOWLEDGE has no information on the question, respond with "I don't know".
        
        If you are asked to write code, you can provide a code based on KNOWLEDGE.       

        Remember, you can only respond based on what you've learned from your brain (KNOWLEDGE). You cannot make up new information.\\
        \midrule
        \textbf{Grading} & 
        You are a grader assessing the relevance of a retrieved document to a current question. \newline 
        Here is the retrieved document: $\{context\}$ \newline
        Here is the user question: $\{question\}$ \newline
        If the document contains keyword(s) or semantic meaning related to the current question, grade it as relevant. \newline
        Give a binary score "yes" or "no" score to indicate whether the document is relevant to the question. \\
        \bottomrule
    \end{tabular}
    \caption{Prompt design for different modes in learner model.} \label{tab:learner_prompt}
\end{table*}

\subsection{Personalized Learner Profile Environment} \label{app:case_study}

AgentTutor offers a personalized learning environment where learners can define their curriculum and receive tailored support across various domains. The system provides multiple dimensions of customization within the teaching session, allowing users to configure the learning experience to match their individual preferences and academic needs.

The configuration options are structured around basic setup and personalized learning profile. The former governs the structural elements of the teaching session, including the number of instructional rounds (which controls session length and depth, adjustable for beginner to deep learning), the question/task type (ranging from general programming to math/algorithm or code implementation problems, influencing the assessment strategy), and the option to display the detailed teaching content generated by the underlying MCTS search process and internal study materials. While the parameters in personalized learner profile will define the tutor's core adaptive behavior: the teaching style (ranging from simple/direct for beginners to detailed/in-depth for advanced learners), the question frequency (from low-frequency for autonomous study to high-frequency for interactive engagement), and the interaction mode (which determines the learners' level of participation, such as highly interactive, passive reception, or a mixed approach, thus influencing the content delivery strategy). The learning curriculum defines the specific learner question or learning topic that initiates the session. For teaching material, learners can also provide their own multimodal resources.

\subsection{Experiments Implementation Details}

\subsubsection{Learner Model} \label{app:learner}
The learner model is designed to simulate confusion, learning, and response processes as a human learner. It is initialized with a knowledge file and the default model (GPT-3.5-turbo). The agent interacts with the knowledge base to load documents and subsequently splits them into a vector store, utilizing the Chroma vector store with the OpenAI embedding model (text-embedding-3-large), integrated into the LangChain framework \cite{langchain}. 

\begin{itemize}
    \item \textbf{Knowledge Base Design.} It initializes with a text file containing the knowledge base and serves as the central repository for managing system knowledge. These documents are segmented into manageable chunks with parameters $chunk\_size=500$ and $chunk\_overlap=200$, ensuring efficient retrieval and processing. \item \textbf{State Graph Design.} The learner operates within a workflow defined by a state graph. This state graph enables the model to manage three main interaction states: confusion, learning, and response, by defining conditional transitions based on the agent's actions and the learner’s knowledge state. For instance, in the confusion mode, if the learner cannot find sufficient information from the knowledge base, it seeks additional information. In the learning mode, the learner can incorporate new knowledge by asking for more information or directly learning from external resources. In the response mode, when adequate information is available, the model generates a response using its stored knowledge. The transitions between these states are controlled by the method, which evaluates the relevance of the retrieved documents to the learner's question. If the documents are deemed relevant, the learner proceeds to generate a response; otherwise, it seeks more information.
    \item \textbf{Learning Process.} The learning process allows the learner to expand its knowledge base by integrating new information. This information can originate from the tutor or URLs that the tutor provides. The new documents are parsed and added to both the knowledge base and the vector store. The learner appends the newly acquired documents to its existing knowledge, splitting them as necessary. If the learner receives external URLs, it attempts to load and parse the content into usable knowledge, retrying if necessary.
    \item \textbf{Response Grading and Generation} The learner's ability to respond to questions is influenced by two major factors. (1) Document Retrieval: This learner agent would retrieve relevant documents based on the current question using its knowledge base. If no documents are found or the retrieved content is not relevant based on the grade method, it may trigger a state change. (2) Response Generation: Once relevant documents are retrieved, the learner agent generates a response, with a carefully structured prompt to ensure the learner's response is based \textbf{solely} on the retrieved knowledge. The grading of the learner's knowledge determines whether the retrieved documents are relevant to the current question. This decision is made based on a binary "yes" or "no" response, indicating whether the information found is applicable to the question. This grading step influences the flow of the state graph, directing the learner either to generate an answer or seek further clarification.
    \item \textbf{Learner Prompt Design.} All the prompt designs for the mentioned modes are provided in Table \ref{tab:learner_prompt}. 

\end{itemize}

\setlength{\cmidrulewidth}{0.1mm} 
\renewcommand{\arraystretch}{1} 
\begin{table*}[ht]
    \centering \small
    \begin{tabular}{ p{2.5cm} c p{11.5cm} }
        \toprule
        \textbf{Metric} & \textbf{Score} & \textbf{Description} \\
        \midrule
        \multirow{5}{2.5cm}{\textbf{Conversational Adaptability}}
        & Criteria & Does the system align teaching strategies with the learner's current cognitive level by effectively responding to user-specific requests? \\
        \cmidrule(lr){2-3}
         & 1 & The system completely fails to recognize or respond to user requests. It shows no understanding of the learner's needs and provides no appropriate feedback, indicating a significant misalignment with the learner's cognitive level. \\
        \cmidrule(lr){2-3} 
        & 2 & The system can partially recognize user requests, but its responses are incomplete or inaccurate. It only captures a fraction of the learner's needs, and the feedback provided does not fully address the learner's cognitive situation. \\
        \cmidrule(lr){2-3}
        & 3 & The system accurately recognizes and responds to most user requests, offering appropriate feedback. It generally aligns with the learner's cognitive level, but there may still be minor discrepancies in understanding complex requests. \\
        \cmidrule(lr){2-3}
        & 4 & The system accurately recognizes and responds to all user requests in a timely and relevant manner. It precisely matches the learner's cognitive level, providing feedback that is well-tailored to the learner's needs. \\
        \cmidrule(lr){2-3}
        & 5 & The system not only accurately recognizes and responds to all user requests but also anticipates the learner's future needs. It proactively provides feedback that exceeds expectations, demonstrating a perfect alignment with the learner's cognitive development. \\
        \midrule
        \multirow{5}{2.5cm}{\textbf{Feedback Quality}}
        & Criteria & Can the system accurately determine answer correctness, estimate proficiency levels, and offer corresponding feedback on learner responses? \\
        \cmidrule(lr){2-3}
        & 1 & The feedback is completely inaccurate or irrelevant. It fails to assess the correctness of the learner's answer, estimate the proficiency level, or provide any useful guidance, making it ineffective for the learning process. \\
        \cmidrule(lr){2-3}
        & 2 & The feedback is partially accurate but lacks depth. It may identify some aspects of the answer's correctness but fails to comprehensively estimate the proficiency level or provide in-depth guidance for improvement. \\
        \cmidrule(lr){2-3}
        & 3 & The feedback is accurate and can effectively assess the learner's response. However, it lacks in-depth guidance on how to improve further, providing only a basic evaluation of the answer and proficiency level. \\
        \cmidrule(lr){2-3}
        & 4 & The feedback is accurate and in-depth. It not only correctly assesses the answer and proficiency level but also provides detailed guidance on how to enhance the learner's performance, facilitating effective learning. \\
        \cmidrule(lr){2-3}
        & 5 & The feedback is not only accurate and in-depth but also stimulates the learner's thinking. It encourages the learner to explore further, promotes self-learning, and significantly contributes to the learner's learning process. \\
        \midrule
        \multirow{5}{2.5cm}{\textbf{Question Difficulty}}
        & Criteria & Do the questions generated by the system have appropriate average difficulty and range, ensuring diversity and depth based on Bloom's Taxonomy and the learner's learning level? \\
        \cmidrule(lr){2-3}
        & 1 & The questions are either too simple or too complex, failing to map to appropriate cognitive levels according to Bloom's Taxonomy. As a result, they cannot effectively evaluate the learner's cognitive level or promote learning. \\
        \cmidrule(lr){2-3}
        & 2 & The questions are of moderate difficulty but lack diversity. They may cover only a limited range of cognitive levels in Bloom's Taxonomy, failing to provide a comprehensive assessment of the learner's abilities. \\
        \cmidrule(lr){2-3}
        & 3 & The questions have appropriate difficulty and diversity, covering different cognitive levels as defined by Bloom's Taxonomy. They offer a balanced assessment of the learner's cognitive skills, facilitating normal learning progress. \\
        \cmidrule(lr){2-3}
        & 4 & The questions are of high difficulty and diversity, spanning a wide range of cognitive levels in Bloom's Taxonomy. They challenge the learner's existing abilities and encourage the development of higher-order thinking skills. \\
        \cmidrule(lr){2-3}
        & 5 & The questions are of extremely high difficulty and diversity, reaching the upper limits of Bloom's Taxonomy. They push the learner's cognitive boundaries, promote deep learning, and foster the development of advanced cognitive skills. \\
        \bottomrule 
    \end{tabular}
    \caption{Prometheus evaluator scoring criteria for three automated metrics, conversational adaptability, feedback quality, and question difficulty, in IESs. }
    \label{tab:automated_metric}
\end{table*}

\subsubsection{Automated Metrics} \label{app:automated_metric}

In evaluating multi-turn interactive processes, we adopt the automated metrics introduced by \citet{jurenka2024towards}, which assesses the extent to which educational systems adhere to best practices in providing appropriate responses, answer feedback, and question formulation. This methodology is crucial for verifying interactive teaching quality. Specifically, we focus on three primary metrics:
\begin{itemize}
    \item \textbf{Conversational Adaptability.} This metric evaluates the system's responsiveness to user-specific requests. It is based on scores from an LLM evaluator that dissects user requests into independent statements and assesses the agent's acknowledgment of these statements in its responses. In our context, this measures whether the educational system's teaching strategies align with the learner's current cognitive level.
    \item \textbf{Feedback Quality.} This metric assesses the quality of the system's feedback on learner responses. It examines whether the system can accurately determine the correctness of learner answers, estimate proficiency levels, and provide corresponding feedback.
    \item \textbf{Question Difficulty.} This metric gauges the average difficulty and range of questions generated by the system to ensure diversity and depth. We categorize question difficulty based on Bloom's Taxonomy, mapping questions to six cognitive levels. The evaluator computes this metric by determining whether the system generates each question according to Bloom's taxonomy, ensuring hierarchical structure and appropriate cognitive load distribution.
    \item \textbf{Automatic Evaluation Details.} Following the methodology outlined in \citet{jiang2024intotheunknown}, we employ the Prometheus model \cite{kim2024prometheus2}, an open-source evaluator designed for assessing long texts based on user-defined criteria. We utilize the prometheus-7B-v2.0 model, setting its default temperature to $1.0$ and top\_p to $0.9$. Due to the model's context window limitations, we remove references and truncate input texts to within $2000$ words during teaching process evaluations, aligning with the practices in \citet{jiang2024intotheunknown}. Table \ref{tab:automated_metric} details the scoring criteria for interactive teaching quality evaluations.

\end{itemize}

\setlength{\cmidrulewidth}{0.1mm} 
\renewcommand{\arraystretch}{1} 
\begin{table*}[htbp]
\centering 
\begin{tabular}{p{2.5cm} l}
\toprule
\textbf{Cognitive Level} & \textbf{Sub-goals} \\
\hline
\textbf{Memory} & 
\begin{tabular}[c]{p{13cm}} 
    - Remember the basic concepts of KNN, such as `distance metric', 'k value', 'classification', and 'regression'. \\
    - Memorize the input and output format of the KNN algorithm. 
\end{tabular} \\
\hline
\textbf{Comprehension} & 
\begin{tabular}[c]{p{13cm}} 
    - Understand how KNN works: using the nearest neighbors to classify or regress data points. \\
    - Understand the impact of selecting the 'k' value on the KNN algorithm's performance. \\
    - Understand how to measure distances between data points using different distance metrics (e.g., Euclidean, Manhattan). 
\end{tabular} \\
\hline
\textbf{Application} & 
\begin{tabular}[c]{p{13cm}} 
    - Apply KNN to simple datasets for classification or regression tasks. \\
    - Choose an appropriate 'k' value and experiment with different values. \\
    - Implement a basic KNN algorithm and validate its output. 
\end{tabular} \\
\hline
\textbf{Analysis} & 
\begin{tabular}[c]{p{13cm}} 
    - Analyze the effect of different 'k' values and distance metrics on the KNN algorithm's results. \\
    - Compare KNN with other machine learning algorithms (e.g., Decision Trees, SVM) and evaluate their pros and cons. \\
    - Identify and analyze the challenges of using KNN in high-dimensional data (e.g., the curse of dimensionality).
\end{tabular} \\
\hline
\textbf{Evaluation} & 
\begin{tabular}[c]{p{13cm}} 
    - Evaluate the performance of a KNN model using metrics such as cross-validation, confusion matrix, accuracy, and recall. \\
    - Tune the KNN model by adjusting the 'k' value and other parameters based on performance evaluations.
\end{tabular} \\
\hline
\textbf{Creation} & 
\begin{tabular}[c]{p{13cm}} 
    - Design optimized versions of the KNN algorithm, considering improvements in computational efficiency (e.g., KD-Tree or Ball-Tree). \\
    - Implement variations of the KNN algorithm, such as Weighted KNN or KNN for regression tasks. \\
    - Explore how to handle large datasets with KNN and overcome related challenges.
\end{tabular} \\
\bottomrule
\end{tabular}
\caption{Learning sub-goals organized by cognitive levels in CDM. We take KNN Algorithm as an illustrative example.} \label{tab:cdm_output}
\end{table*}

\end{document}